\documentclass[reprint,aps,pra,showpacs,floatfix,twocolumn]{revtex4-1}
\usepackage{graphicx}
\usepackage{dsfont}
\usepackage{epstopdf}

\newcommand{\be}{\begin{equation}}
\newcommand{\ee}{\end{equation}}

\begin{document}

\title{Perfect state transfer by means of discrete-time quantum walk search algorithms on highly symmetric graphs}

\author{M. \v Stefa\v n\'ak\email[correspondence to:]{martin.stefanak@fjfi.cvut.cz}}
\affiliation{Department of Physics, Faculty of Nuclear Sciences and Physical Engineering, Czech Technical University in Prague, B\v
rehov\'a 7, 115 19 Praha 1 - Star\'e M\v{e}sto, Czech Republic}

\author{S. Skoup\'y}
\affiliation{Department of Physics, Faculty of Nuclear Sciences and Physical Engineering, Czech Technical University in Prague, B\v
rehov\'a 7, 115 19 Praha 1 - Star\'e M\v{e}sto, Czech Republic}


\date{\today}

\begin{abstract}
Perfect state transfer between two marked vertices of a graph by means of discrete-time quantum walk is analyzed. We consider the quantum walk search algorithm with two marked vertices, sender and receiver. It is shown by explicit calculation that for the coined quantum walks on star graph and complete graph with self-loops perfect state transfer between the sender and receiver vertex is achieved for arbitrary number of vertices $N$ in $O(\sqrt{N})$ steps of the walk. Finally, we show that Szegedy's walk with queries on complete graph allows for state transfer with unit fidelity in the limit of large $N$.
\end{abstract}

\maketitle


\section{Introduction}

Quantum walks \cite{adz} have emerged as quantum analogues of a classical random walk on a discrete lattice or a graph. Both discrete-time \cite{meyer} and continuous time \cite{fg} quantum walks were proposed. Soon, the potential of quantum walks in quantum information processing was identified \cite{qw:aaku}. In fact, it was found that both continuous-time \cite{childs} and discrete-time \cite{Lovett} quantum walks are universal models of quantum computation.

One of the most prominent application of quantum walks in quantum information processing is the spatial search of the unsorted database of $N$ items represented by a graph with a marked vertex. Marking the vertex corresponds to different dynamics on that node, i.e. different coin operator in the discrete-time quantum walk or different on-site energy in the continuous-time quantum walk. Discrete-time quantum walk search algorithm was shown to be optimal for hypercube \cite{skw} and for lattices \cite{ambainis} of dimensions $d$ greater than 2, i.e. it finds the marked node after $O(\sqrt{N})$ steps of the walk. Continuous-time quantum walk was shown to be optimal \cite{childs:search} for search on the complete graph, hypercube and lattices with $d>4$. Moreover, including the coin degree of freedom the continuous-time quantum walk search is optimal for lattices with $d>2$ \cite{childs:cont:coin}. Later it was found that high symmetry or connectivity of the graph is in fact not required for the optimal runtime of the continuous-time quantum walk search algorithm \cite{janmark:search,novo:dim:red,meyer:search}. In fact, Chakraborty et al. \cite{shantanav:search} have shown that continuous-time quantum walk search algorithm is optimal for almost all graphs. Another variant of discrete-time coinless quantum walk capable of optimal search was proposed by Szegedy \cite{szegedy}. Szegedy's walk on complete graph finds the marked vertex with probability $1/2$. Recently, Santos \cite{santos} have found that adding queries to the Szegedy's walk on the complete graph increases the probability of finding the marked vertex to 1 in the limit of large $N$.

Another promising application of quantum walks is the perfect state transfer between two vertices of a graph or a lattice. There exist two different approaches to the problem. In the first one defines dynamics at each individual vertex in order to achieve state transfer between two selected vertices. This approach was pursued by Kurzynski and Wojcik \cite{wojcik:qw:pst}, who have designed the local coin operators to achieve perfect state transfer with discrete-time quantum walk on a circle. The method of \cite{wojcik:qw:pst} is essentially the discrete-time variant of the engineered coupling protocol \cite{christandl} in spin chains. In a similar way, Zhan et al. \cite{zhan:qw:pst} have designed paths using local coin operators of discrete time quantum walk, either identity matrices or tensor product of Pauli $\sigma_x$, which leads to state transfer on a square lattice. Yalcinkaya and Gedik \cite{gedik:qw:pst} have analyzed the state transfer on a circle with fixed coin operator. They have shown that only identity or Pauli $\sigma_x$ achieves state transfer with unit fidelity over arbitrary distance, while Hadamard operator or other mixing coins allow for perfect state transfer over finite distances only. In these models \cite{wojcik:qw:pst,zhan:qw:pst,gedik:qw:pst} the transfer of the internal coin state is also possible. Second approach, where one modifies the dynamics only at vertices which want to communicate the quantum state, was proposed by Hein and Tanner \cite{hein:wave:com}. The authors have considered discrete-time quantum walk search algorithm on a lattice with two marked vertices, sender and receiver, and showed that initializing the algorithm on the sender vertex the walk will reach the receiver vertex with high probability. In this scenario only the transfer of particle from one vertex to the other is considered, instead of the transfer of arbitrary internal coin state. For finite graphs, especially cycles and their variants, this approach was analyzed by \cite{kendon:qw:pst,barr:pst} in both discrete-time and continuous-time models. More recently, Chakraborty et al. \cite{shantanav:search} have shown that in the continuous-time quantum walk scenario it is possible to achieve perfect state transfer for almost any graph in the limit of large size of the graph $N$.

In the present paper we follow the idea of Hein and Tanner \cite{hein:wave:com} for perfect state transfer by means of discrete-time quantum walk on highly symmetric graphs. We focus on such graphs where the discrete-time quantum walk search algorithm succeeds in finding the marked vertex with certainty, namely the star graph and complete graph with self-loops \cite{ambainis,wong}. We also consider Szegedy's walk with queries on the complete graph \cite{santos} where unit success probability is reached in the limit of large size of the graph $N$. We explicitly show that the algorithms are capable of state transfer between the sender and the receiver vertices in $O(\sqrt{N})$ steps. The method is analogous to the analysis of the search algorithms on the corresponding graphs \cite{ambainis,reitzner:search,wong,santos}. Namely, we determine the invariant subspace of the evolution operator of the walk which includes the sender and the receiver states. Since the distance between the sender and the receiver vertices in the models discussed in the present paper is independent of the size of the graph $N$ the dimension of the invariant subspace is also independent of $N$. Similar dimensional reduction due to the high symmetry of the graph \cite{krovi} was also applied previously in analysis of anomaly identification on star graphs \cite{feldman:struct,hillery:probe} and continuous-time quantum walk search algorithms \cite{novo:dim:red}. In particular, the invariant subspace has dimension 3 for the star graph, 5 for the complete graph with self-loops and 7 for the Szegedy's walk with queries on the complete graph. This fact greatly reduces the complexity of the problem. Indeed, we only have to deal with the effective evolution operator which is a fixed size matrix with matrix elements depending on the size of the graph $N$. For star graph and complete graph with self-loops the effective evolution operator can be diagonalized analytically and the problem of state transfer can be solved exactly. We show that for both graphs the quantum walk achieves perfect state transfer, i.e. the particle is transferred with unit probability, for arbitrary size of the graph $N$. In the case of the Szegedy's walk with queries on complete graph we show that the particle is transferred with unit probability in the limit of large $N$.

Our manuscript is organized as follows: In Section~\ref{sec2} we analyze the perfect state transfer in the coined quantum walk on the star graph. Section~\ref{sec3} is devoted to perfect state transfer in the coined quantum walk on the complete graph with self-loops. Finally, state transfer in the Szegedy's walk with queries on the complete graph is discussed in Section~\ref{sec4}. We summarize our results in the conclusions of Section~\ref{sec5}.


\section{Star graph}
\label{sec2}

Let us begin with the state transfer between two vertices of a star graph by means of a discrete-time quantum walk. Discrete-time quantum walk search algorithm on the star graph is exactly equivalent to the Grover search algorithm \cite{grover:search}, hence, it finds the marked vertex with unit probability. We show by explicit calculation that the algorithm also achieves perfect state transfer.

Star graph consists of a central vertex labeled as 0 which is connected to $N$ external vertices with labels 1 to $N$. Discrete-time quantum walk on the star graph can be defined as a scattering walk \cite{feldman:struct,hillery:probe} or as the usual coined quantum walk. Both models are equivalent \cite{equiv:coin:scat,equiv:coin:scat:2}, and since the coined walk will be used in the following Section~\ref{sec3} we pursue this approach. We consider a quantum walk where the particle jumps from the external vertices to the central vertex and back. The position space is spanned by the vectors $|j\rangle_p$, with $j=0,\ldots, N$, corresponding to the particle being at the vertex $j$. The coin space has to be defined separately for the external vertices and for the central vertex. At the external nodes the coin space is one-dimensional, since the particle can jump only to the central vertex $0$. We denote the coin state as $|0\rangle_c$. At the central node the coin space has a dimension $N$, as the particle is allowed to jump to any external vertex $j$, with $j=1,\ldots, N$. We denote the corresponding coin states as $|j\rangle_c$. The complete Hilbert space of the discrete-time quantum walk on the star graph is therefore spanned by vectors
\begin{eqnarray}
\nonumber |j\rangle_p\otimes|0\rangle_c & \equiv & |j,0\rangle, \\
\nonumber |0\rangle_p\otimes|j\rangle_c & \equiv & |0,j\rangle,
\end{eqnarray}
where $j$ runs from 1 to $N$. The first index corresponds to the vertex and the second index corresponds to the coin state.

The evolution operator of a single step of the walk can be written as a product of the step operator $S$ and the coin operator $C$
\begin{equation}
\label{evol:star}
U = S\cdot C.
\end{equation}
The walk describes the particle hopping between the external vertices and the central node. Hence, the step operator is given by
$$
S = \sum_{j=1}^N \left(|j,0\rangle\langle 0,j| + |0,j\rangle\langle j,0|\right).
$$
Let us now turn to the coin operator. At the external nodes, where the coin space is one-dimensional, we choose the coin operator to act as identity. However, for the sake of state transfer, we have two marked vertices $s$ (sender) and $r$ (receiver), where the coin acts as a phase shift of $\pi$. At the central node the states $|j\rangle_c$ form an $N$-dimensional space, and we choose the coin operator to act there as the Grover diffusion operator
\begin{equation}
\label{grover:op}
G = 2|\psi_S\rangle_c\langle\psi_S| - I_N,
\end{equation}
where $|\psi_S\rangle_c$ denotes the symmetric superposition of all basis states $|j\rangle_c$
\begin{equation}
\label{eqws}
|\psi_S\rangle_c = \frac{1}{\sqrt{N}}\sum_{j=1}^N |j\rangle_c,
\end{equation}
and $I_N$ is the identity operator on the Hilbert space of dimension $N$. Hence, the coin operator is defined as
$$
C = \left(I_N - 2|s\rangle_p\langle s| - 2|r\rangle_p\langle r|\right)\otimes |0\rangle_c\langle 0| + |0\rangle_p\langle 0|\otimes G.
$$
After some algebra we find that the evolution operator (\ref{evol:star}) can be re-written as
\begin{eqnarray}
\label{U:star}
\nonumber U & = & \sum_{j=1}^N |0,j\rangle\langle j,0| - 2|0,s\rangle\langle s,0| - 2|0,r\rangle\langle r,0| + \\
& & + \frac{2}{N}\sum_{i,j=1}^N |i,0\rangle\langle 0,j| - \sum_{j=1}^N |j,0\rangle\langle 0,j|.
\end{eqnarray}
We start the walk in the sender vertex, i.e. the initial state is
$$
|\psi(0)\rangle = |s,0\rangle.
$$
The state of the walk after $t$ steps is given by
$$
|\psi(t)\rangle = U^t |\psi(0)\rangle.
$$
We will show that after $O(\sqrt{N})$ steps the particle will be on the receiver vertex, i.e. in the state $|r,0\rangle$. Clearly, the walk is bipartite, since in the odd steps the particle is at the central node and in the even steps it is at the external nodes. Since we want to analyze the possibility of state transfer between two external nodes $s$ and $r$ we focus only on the square of the evolution operator. From the expression (\ref{U:star}) the action of $U^2$ on the states $|j,0\rangle$ is then easily found to be
\begin{eqnarray}
\label{U2:star}
\nonumber U^2 |j,0\rangle & = & \frac{2}{N}\sum_{i\neq j}|i,0\rangle - \left(1-\frac{2}{N}\right)|j,0\rangle,\quad j\neq s,r\\
\nonumber U^2 |s,0\rangle & = & -\frac{2}{N}\sum_{i\neq s}|i,0\rangle + \left(1-\frac{2}{N}\right)|s,0\rangle,\\
U^2 |r,0\rangle & = & -\frac{2}{N}\sum_{i\neq r}|i,0\rangle + \left(1-\frac{2}{N}\right)|r,0\rangle .
\end{eqnarray}
Using these expressions one shows that the following three orthogonal states
\begin{eqnarray}
\label{alpha:star}
\nonumber |\alpha_1\rangle & = & |s,0\rangle, \\
\nonumber |\alpha_2\rangle & = & |r,0\rangle, \\
|\alpha_3\rangle & = & \frac{1}{\sqrt{N-2}}\sum_{j\neq s, r}|j,0\rangle,
\end{eqnarray}
form an invariant subspace with respect to $U^2$. Indeed, from (\ref{U2:star}) we find
\begin{eqnarray}
\nonumber U^2|\alpha_1\rangle & = & \left(1-\frac{2}{N}\right)|\alpha_1\rangle - \frac{2}{N}|\alpha_2\rangle - \frac{2\sqrt{N-2}}{N}|\alpha_3\rangle,\\
\nonumber U^2|\alpha_2\rangle & = & - \frac{2}{N}|\alpha_1\rangle + \left(1-\frac{2}{N}\right)|\alpha_2\rangle  - \frac{2\sqrt{N-2}}{N}|\alpha_3\rangle,\\
\nonumber U^2|\alpha_3\rangle & = & \frac{2\sqrt{N-2}}{N}(|\alpha_1\rangle + |\alpha_2\rangle)  - \left(1-\frac{4}{N}\right)|\alpha_3\rangle.
\end{eqnarray}
Hence, the time evolution of the walk for the fixed initial state $|\alpha_1\rangle$ is described by the effective evolution operator $U_{eff}$, which is in the $|\alpha_i\rangle$ basis (\ref{alpha:star}) given by the following 3x3 matrix
$$
U_{eff} = \left(
              \begin{array}{ccc}
                1-\frac{2}{N} & -\frac{2}{N} & \frac{2\sqrt{N-2}}{N} \\
                -\frac{2}{N} & 1-\frac{2}{N} & \frac{2\sqrt{N-2}}{N} \\
                -\frac{2\sqrt{N-2}}{N} & -\frac{2\sqrt{N-2}}{N} & 1-\frac{4}{N} \\
              \end{array}
            \right).
$$
Diagonalization of $U_{eff}$ is straightforward. We find that it has an eigenvector
\begin{equation}
\label{star:evec:0}
|\chi_0\rangle = \frac{1}{\sqrt{2}}\left(|\alpha_1\rangle - |\alpha_2\rangle\right),
\end{equation}
corresponding to the eigenvalue $\lambda = 1$. The remaining two eigenvectors have the form
\begin{equation}
\label{star:evec:pm}
|\chi^\pm\rangle = \frac{1}{2}\left(|\alpha_1\rangle + |\alpha_2\rangle\right) \pm \frac{i}{\sqrt{2}}|\alpha_3\rangle.
\end{equation}
They correspond to a pair of conjugated eigenvalues
$$
\lambda^\pm = e^{\pm i\omega},
$$
where the phase $\omega$ is given by
\begin{equation}
\label{omega:star}
\omega = \arccos{\left(\frac{N-4}{N}\right)}.
\end{equation}
Let us now analyze the evolution of the initial state $|\alpha_1\rangle$ under the effective evolution operator $U_{eff}$. We find that the initial condition $|\alpha_1\rangle$ and the desired target state $|\alpha_2\rangle$ can be decomposed into the eigenbasis of $U_{eff}$ as
\begin{eqnarray}
\nonumber |\alpha_1\rangle & = & \frac{1}{\sqrt{2}}|\chi_0\rangle + \frac{1}{2}\left(|\chi^+\rangle + |\chi^-\rangle\right),\\
\nonumber |\alpha_2\rangle & = & -\frac{1}{\sqrt{2}}|\chi_0\rangle + \frac{1}{2}\left(|\chi^+\rangle + |\chi^-\rangle\right).
\end{eqnarray}
After $t$ applications of the effective evolution operator $U_{eff}$, i.e. after $2t$ steps of the walk, we obtain
\begin{equation}
\label{state:star}
|\psi(2t)\rangle = \frac{1}{\sqrt{2}}|\chi_0\rangle + \frac{e^{i\omega t}}{2}\left(|\chi^+\rangle + e^{-2i\omega t}|\chi^-\rangle\right).
\end{equation}
For $\omega t = \pi$ the state reduces to $-|\alpha_2\rangle$, i.e. the receiver state up to an irrelevant global phase factor. We conclude that the walk achieves (almost) perfect state transfer between the sender and receiver vertices after $T$ steps, provided that we choose $T$ as the closest integer to $2\pi/\omega$, i.e.
\begin{equation}
\label{T:star}
T \approx \frac{2\pi}{\arccos{\left(\frac{N-4}{N}\right)}}.
\end{equation}
With the Taylor expansion we find that the number of steps required for the state transfer scales with the size of the star graph according to
$$
T \sim \frac{\pi}{\sqrt{2}}\sqrt{N} + O(N^{-\frac{1}{2}}).
$$

\begin{figure}[h]
\begin{center}
\includegraphics[width=0.45\textwidth]{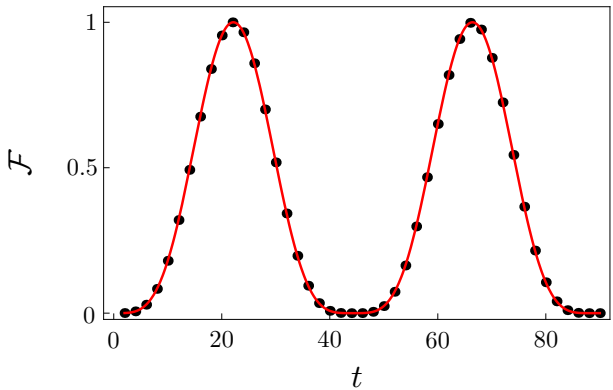}\hfill
\end{center}
\caption{Fidelity between the state of the walk (\ref{state:star}) and the target state $|\alpha_2\rangle$ for the walk on the star graph as a function of the number of steps $t$. The black dots correspond to the numerical simulation and the red line is given by (\ref{fidelity:star}). Fidelity is plotted only at even number of steps, since it vanishes when $t$ is odd. We have considered the star graph with $N=100$ external vertices. The first maximum of fidelity is reached after 22 steps, in accordance with (\ref{T:star}).}
\label{fig1}
\end{figure}

For illustration we display in Figure~\ref{fig1} the fidelity between the state of the walk (\ref{state:star}) and the target state $|\alpha_2\rangle$ as a function of the number of steps. From (\ref{state:star}) we find that it is given by
\begin{equation}
\label{fidelity:star}
{\cal F}(2t) = \left|\langle\psi(2t)|\alpha_2\rangle\right|^2 = \sin^4\left(\frac{\omega t}{2}\right).
\end{equation}
Note that for odd time steps the fidelity is zero since the walk is bipartite. In Figure~\ref{fig1} the number of vertices of the star graph was chosen as $N=100$. As follows from (\ref{T:star}) the first maximum of the fidelity is reached after 22 steps of the walk.

\section{Complete graph with self-loops}
\label{sec3}

Let us now turn to the state transfer on the complete graph of $N$ vertices with additional self-loop on each vertex. The reason we consider the additional self-loops is that the discrete-time quantum walk search algorithm on the complete graph does not find the marked vertex with unit probability. Nevertheless, it was shown \cite{ambainis,wong} that adding self-loops makes two steps of the discrete-time quantum walk equivalent to the Grover search algorithm and increases probability of finding the marked vertex to one. In the following we show explicitly that the algorithm achieves state transfer with unit fidelity independent of the size of the graph.

The Hilbert space of the walk is given by
$$
{\mathcal H} = {\mathcal H}_P\otimes {\mathcal H}_C,
$$
where both position space and coin space have dimension $N$. We denote the basis vectors of ${\mathcal H}_P$ as  $|1\rangle_p,\ldots ,|N\rangle_p$. Similarly, the basis vectors of ${\mathcal H}_C$ are denoted as  $|1\rangle_c,\ldots ,|N\rangle_c$. The basis of $\mathcal H$ is then formed by the vectors $|i\rangle_p\otimes|j\rangle_c\equiv|i,j\rangle$, where the first index corresponds to the position (vertex), and the second index corresponds to the coin state.

The evolution operator of the walk is given by the product of the step operator and the coin operator
$$
U = S\cdot C.
$$
The step operator reads
$$
S = \sum_{i,j=1}^N|j,i\rangle\langle i,j|.
$$
As for the coin operator, we choose it to act as the Grover operator (\ref{grover:op}) on all non-marked vertices, with an additional phase shift of $\pi$ on the marked vertices $s$ and $r$. Hence, $C$ can be written as
$$
C = \left(I_N - 2|s\rangle_p\langle s| - 2|r\rangle_p\langle r|\right)\otimes G,
$$
where $G$ is given in (\ref{grover:op}).

Concerning the initial state of the walk, we choose the particle to be localized on the sender vertex $s$ with the equal weight superposition of all coin states (\ref{eqws}), i.e.
$$
|\psi(0)\rangle = |s\rangle_p\otimes|\psi_S\rangle_c = \frac{1}{\sqrt{N}}\sum_{j=1}^N |s,j\rangle.
$$
We again denote this state as $|\alpha_1\rangle$ since it will be the first basis vector of the invariant subspace. We now show that after $O(\sqrt{N})$ steps of the walk the particle will be in the state
$$
|\alpha_2\rangle = |r\rangle_p\otimes|\psi_S\rangle_c = \frac{1}{\sqrt{N}}\sum_{j=1}^N |r,j\rangle,
$$
i.e. localized on the receiver vertex $r$. Similarly like for the star graph, it is sufficient to consider $U^2$, since \cite{ambainis,wong} have shown that two steps of the walk are equivalent to one iteration of the Grover search algorithm on the position Hilbert space ${\mathcal H}_P$. First, let us determine the invariant subspace of $U^2$ which includes $|\alpha_{1,2}\rangle$. Simple algebra reveals that the following four orthonormal vectors
\begin{eqnarray}
\label{alpha'}
\nonumber |\alpha_3'\rangle & = & \frac{1}{\sqrt{2(N-2)}}\sum _{i\neq s,r} \left(|i,s\rangle + |i,r\rangle\right),\\
 |\alpha_4'\rangle & = & \frac{1}{N-2}\sum_{i,j\neq s,r}|i,j\rangle, \\
\nonumber |\alpha_5'\rangle & = &\sqrt{\frac{2}{N-2}}|\alpha_1\rangle - \sqrt{\frac{N}{2 (N-2)}}(|s,s\rangle + |s,r\rangle),\\
\nonumber |\alpha_6'\rangle & = &\sqrt{\frac{2}{N-2}}|\alpha_2\rangle - \sqrt{\frac{N}{2 (N-2)}}(|r,s\rangle + |r,r\rangle),
\end{eqnarray}
complement $|\alpha_{1,2}\rangle$ to the invariant subspace of $U^2$. However, we can reduce the dimension of the invariant subspace further from 6 to 5. Indeed, one can show that $U^2$ has an eigenvector
$$
|\chi\rangle = \frac{1}{\sqrt{N}}|\alpha_3'\rangle + \sqrt{\frac{N-2}{2N}}|\alpha_4'\rangle + \frac{1}{2}|\alpha_5'\rangle + \frac{1}{2}|\alpha_6'\rangle,
$$
corresponding to the eigenvalue 1, which is orthogonal to $|\alpha_{1,2}\rangle$. Hence, $|\chi\rangle$ is also orthogonal to $U^2|\alpha_{1,2}\rangle$, and thus it can be subtracted from the invariant subspace. The orthogonal complement of $|\chi\rangle$ in the subspace spanned by vectors (\ref{alpha'}) then completes $|\alpha_{1,2}\rangle$ to the invariant subspace of of $U^2$. We choose the orthonormal basis as
\begin{widetext}
\begin{eqnarray}
\nonumber |\alpha_3\rangle & = & \sqrt{\frac{N-2}{N}}|\alpha_3'\rangle  - \sqrt{\frac{2}{N}} |\alpha_4'\rangle  = \frac{1}{\sqrt{2N}}\sum_{i=1}^N (|i,s\rangle+|i,r\rangle) - \frac{\sqrt{2}}{(N-2)\sqrt{N}}\sum_{i,j\neq s,r}|i,j\rangle,\\
\nonumber |\alpha_4\rangle & = & \frac{1}{\sqrt{2}}|\alpha_5'\rangle  - \frac{1}{\sqrt{2}}|\alpha_6'\rangle  = \frac{1}{\sqrt{N(N-2)}}\sum_{j\neq s,r}(|s,j\rangle - |r,j\rangle) + \sqrt{\frac{N-2}{4N}}\left(|r,r\rangle + |r,s\rangle - |s,s\rangle - |s,s\rangle\right),\\
\nonumber |\alpha_5\rangle & = &  \frac{1}{\sqrt{N}}|\alpha_3'\rangle + \sqrt{\frac{N-2}{2N}}|\alpha_4'\rangle - \frac{1}{2}|\alpha_5'\rangle - \frac{1}{2}|\alpha_6'\rangle \\
\nonumber & = & \frac{1}{\sqrt{2N(N-2)}}\left(\sum_{i,j\neq s,r} |i,j\rangle + \sum_{i\neq s,r} (|i,s\rangle + |i,r\rangle )  - \sum_{j\neq s,r} (|s,j\rangle + |r,j\rangle ) \right) + \\
\nonumber & & + \sqrt{\frac{N-2}{8N}}\left(|s,r\rangle + |s,s\rangle + |r,s\rangle + |r,r\rangle\right).
\end{eqnarray}
The effective evolution operator in the $|\alpha_i\rangle$ basis is given by the matrix
$$
U_{eff} = \left(
\begin{array}{ccccc}
 \frac{(N-4) (N-2)}{N^2} & -\frac{2(N-4)}{N^2} & \frac{4 \sqrt{2} (N-2)}{N^2} & -\frac{2 \sqrt{N-2}}{N} & \frac{2 \sqrt{2} (N-4) \sqrt{N-2}}{N^2} \\
 -\frac{2(N-4)}{N^2} & \frac{(N-4) (N-2)}{N^2} & \frac{4 \sqrt{2} (N-2)}{N^2} & \frac{2 \sqrt{N-2}}{N} & \frac{2 \sqrt{2} (N-4) \sqrt{N-2}}{N^2} \\
 \frac{4 \sqrt{2} (N-2)}{N^2} & \frac{4 \sqrt{2} (N-2)}{N^2} & \frac{(N-4)^2}{N^2} & 0 & -\frac{4 (N-4) \sqrt{N-2}}{N^2} \\
 \frac{2 \sqrt{N-2}}{N} & -\frac{2 \sqrt{N-2}}{N} & 0 & \frac{N-4}{N} & 0 \\
 -\frac{2 \sqrt{2} (N-4) \sqrt{N-2}}{N^2} & -\frac{2 \sqrt{2} (N-4) \sqrt{N-2}}{N^2} & \frac{4 (N-4) \sqrt{N-2}}{N^2} & 0 & \frac{N^2-16 N+32}{N^2} \\
\end{array}
\right)
$$
\end{widetext}
We find that the spectrum of $U_{eff}$ consists of eigenvalues
\begin{eqnarray}
\label{spectrum:2}
\nonumber \lambda_0 & = & 1,\\
\nonumber \lambda_{1}^\pm & = & e^{\pm i\omega},\\
 \lambda_{2}^\pm & = & e^{\pm 2i\omega},
\end{eqnarray}
where the phase $\omega$ is given in (\ref{omega:star}). The corresponding eigenvectors are found to be
\begin{eqnarray}
\label{eb:cl}
\nonumber |\chi_0\rangle & = &  \frac{1}{2}|\alpha_1\rangle + \frac{1}{2}|\alpha_2\rangle +\frac{1}{\sqrt{2}}|\alpha_3\rangle,\\
 |\chi_1^\pm\rangle & = & \frac{1}{2}|\alpha_1\rangle - \frac{1}{2}|\alpha_2\rangle \mp \frac{i}{\sqrt{2}}|\alpha_4\rangle,\\
\nonumber |\chi_2^\pm\rangle & = & \frac{1}{2\sqrt{2}}|\alpha_1\rangle + \frac{1}{2\sqrt{2}}|\alpha_2\rangle -\frac{1}{2}|\alpha_3\rangle \pm \frac{i}{\sqrt{2}}|\alpha_5\rangle.
\end{eqnarray}
The initial state of the walk $|\alpha_1\rangle$ and the desired target state $|\alpha_2\rangle$ are decomposed into the eigenbasis (\ref{eb:cl}) of effective evolution operator according to
\begin{eqnarray}
\nonumber |\alpha_1\rangle & = & \frac{1}{2}|\chi_0\rangle + \frac{1}{2}\left(|\chi_1^+\rangle + |\chi_1^-\rangle\right) + \frac{1}{2\sqrt{2}}\left(|\chi_2^+\rangle + |\chi_2^-\rangle\right),\\
\nonumber |\alpha_2\rangle & = & \frac{1}{2}|\chi_0\rangle - \frac{1}{2}\left(|\chi_1^+\rangle + |\chi_1^-\rangle\right) + \frac{1}{2\sqrt{2}}\left(|\chi_2^+\rangle + |\chi_2^-\rangle\right)
\end{eqnarray}
After $2t$ steps of the walk the state can be written as
\begin{eqnarray}
\label{state:complete}
\nonumber |\psi(2t)\rangle & = & U_{eff}^t|\alpha_1\rangle \\
\nonumber & = & \frac{1}{2}|\chi_0\rangle + \frac{e^{i\omega t}}{2}\left(|\chi_1^+\rangle + e^{-2i\omega t}|\chi_1^-\rangle\right) + \\
& & + \frac{e^{2i\omega t}}{2\sqrt{2}}\left(|\chi_2^+\rangle + e^{-4i\omega t} |\chi_2^-\rangle\right).
\end{eqnarray}
We find that for $\omega t = \pi$ the state reduces to the desired target state $|\alpha_2\rangle$. Hence, to achieve perfect state transfer we have to choose the number of steps $T$ as the closest integer to $\frac{2\pi}{\omega}$, which is exactly the same as for the star graph (\ref{T:star}). We note that the perfect state transfer in this model is possible for arbitrary $N$ thanks to the perfect matching of the spectrum (\ref{spectrum:2}), i.e. the fact that the phases of eigenvalues $\lambda_2^\pm$ are exactly twice the phases of the eigenvalues of $\lambda_1^\pm$.

\begin{figure}[h]
\begin{center}
\includegraphics[width=0.45\textwidth]{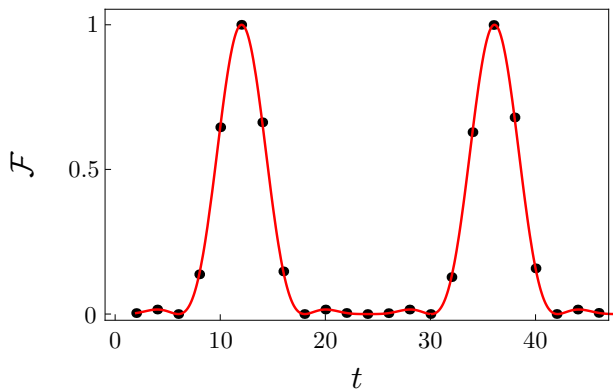}\hfill
\end{center}
\caption{Fidelity between the state of the walk (\ref{state:complete}) and the target state $|\alpha_2\rangle$ for the walk on the complete graph with self-loops as a function of the number of steps $t$. The black dots correspond to the numerical simulation and the red line is given by (\ref{fidelity:complete}). Fidelity is plotted only at even number of steps. We have considered the complete graph with self-loops with $N=30$ vertices. The first maximum of fidelity is reached after 12 steps, in accordance with (\ref{T:star}).}
\label{fig2}
\end{figure}

For illustration we display in Figure~\ref{fig2} the fidelity between the state of the walk (\ref{state:complete}) and the target state $|\alpha_2\rangle$ as a function of the number of steps, which is given by
\begin{equation}
\label{fidelity:complete}
{\cal F}(2t) = \left|\langle\psi(2t)|\alpha_2\rangle\right|^2 = \cos^2(\omega t)\sin^4\left(\frac{\omega t}{2}\right).
\end{equation}
In comparison to the result for the star graph (\ref{fidelity:star}) we find that there is an additional modulation with $\cos^2(\omega t)$ arising from the eigenvectors $|\chi_2^\pm\rangle$ that oscillate at double frequency. In Figure~\ref{fig2} the number of vertices was chosen as $N=30$. The first maximum of fidelity is reached after 12 steps of the walk, in agreement with the analytical prediction of (\ref{T:star}).

\section{Szegedy's walk with queries on the complete graph}
\label{sec4}

Finally, let us consider the state transfer in Szegedy's walk, which is a coinless discrete-time quantum walk model driven by reflection operators in a bipartite graph \cite{portugal:szegedy}. In the original proposal of the Szegedy's walk \cite{szegedy} the search algorithm finds the marked vertex of the complete graph with probability $\frac{1}{2}$. However, Santos \cite{santos} have shown that adding phase shifts of $\pi$ on the marked vertices (i.e. queries), increases the success probability to one in the limit of large number of vertices $N$. Therefore, we consider the Szegedy's walk with queries on the complete graph with two marked vertices $s$ and $r$. We show that in the limit of large $N$ the walk achieves perfect state transfer between the sender and the receiver.

Let us briefly review the definition of the Szegedy's walk \cite{szegedy} on the graph $G(X,E)$, where $X = \left\{1,\ldots, N\right\}$ is the set of vertices and $E$ is the set of edges. We turn it to bipartite graph of $N+N$ vertices, i.e. duplicate the graph G, remove all edges in the original graph and its copy, and add edges between the vertices in the two sets as in the original graph. The Hilbert space of the Szegedy's walk is given by tensor product of two $N$-dimensional Hilbert spaces ${\mathcal H}_N$
$$
{\mathcal H} = {\mathcal H}_N \otimes {\mathcal H}_N,
$$
corresponding to the vertices of the original graph and its copy. We denote the vectors of computational basis of ${\mathcal H}$ as
$$
|i\rangle\otimes |j\rangle \equiv |i,j\rangle,\quad i,j = 1,\ldots , N,
$$
where the first index corresponds to the vertex of the original graph and the second index denotes the vertex in the copy. Szegedy's walk \cite{szegedy} is driven by reflections around subspaces generated by vectors $|\Phi_i\rangle$ and $|\Psi_j\rangle$
\begin{eqnarray}
\nonumber {\cal R}_A & = & 2\sum_{i=1}^N|\Phi_i\rangle\langle\Phi_i| - I_{N^2}, \\
\nonumber {\cal R}_B & = & 2\sum_{j=1}^N|\Psi_j\rangle\langle\Psi_j| - I_{N^2},
\end{eqnarray}
which are defined as
\begin{eqnarray}
\label{szegedy:vec}
\nonumber |\Phi_i\rangle & = & |i\rangle\otimes \left(\sum_{j}\sqrt{p_{ij}}|j\rangle\right),\\
|\Psi_j\rangle & = & \left(\sum_{i}\sqrt{p_{ij}}|i\rangle\right)\otimes |j\rangle.
\end{eqnarray}
Here $p_{ij}$ denotes components of a stochastic matrix associated to the graph $G$. We consider $G$ to be the complete graph and for simplicity take the stochastic matrix as
$$
p_{ij} = \frac{1}{N-1}\left(1-\delta_{ij}\right).
$$
Hence, in our model the vectors (\ref{szegedy:vec}) are given by
\begin{eqnarray}
\nonumber  |\Phi_i\rangle & = & \frac{1}{\sqrt{N-1}}\sum_{j\neq i}|i,j\rangle, \\
\nonumber |\Psi_j\rangle & = & \frac{1}{\sqrt{N-1}}\sum_{i\neq j}|i,j\rangle.
\end{eqnarray}
Santos \cite{santos} has extended the evolution of the Szegedy's walk with queries, i.e phase shift of $\pi$ on the marked vertices. Since we have two marked vertices $s$ and $r$, the action of the queries is described by the following operator
$$
{\cal R}_M  =  \left(I_N-2|s\rangle\langle s| - 2|r\rangle\langle r|\right)\otimes I_N.
$$
The complete evolution operator of the Szegedy's walk with queries is then given by \cite{santos}
\begin{equation}
\label{evol:szegedy}
U = {\cal R}_B{\cal R}_A{\cal R}_M.
\end{equation}
We show that for large $N$, starting the walk in the state
$$
|\alpha_1\rangle = |\Phi_s\rangle = \frac{1}{\sqrt{N-1}}\sum_{j\neq s}|s,j\rangle,
$$
and performing $O(\sqrt{N})$ steps we will obtain with high probability the state
$$
|\alpha_2\rangle = |\Phi_r\rangle = \frac{1}{\sqrt{N-1}}\sum_{j\neq r}|r,j\rangle.
$$
Notice that in the first vector the first index is $s$, while in the second vector the first index $r$. In this sense, we achieve the state transfer from the vertex $s$ to vertex $r$.

First, we determine the invariant subspace which includes the initial and the final states $|\alpha_1\rangle$ and $|\alpha_2\rangle$. Using the definition of the evolution operator (\ref{evol:szegedy}) we find that the invariant subspace includes five additional orthonormal vectors
\begin{eqnarray}
\nonumber |\alpha_3\rangle & = & \frac{1}{\sqrt{(N-2)(N-3)}}\sum\limits_{\begin{array}{c}
                                                               i,j\neq s,r \\
                                                               i\neq j
                                                             \end{array}
}|i,j\rangle,\\
\nonumber |\alpha_4\rangle & = & \frac{1}{\sqrt{(N-1)(N-2)}}\sum_{j\neq s,r}|s,j\rangle - \sqrt{\frac{N-2}{N-1}}|s,r\rangle,\\
\nonumber |\alpha_5\rangle & = & \frac{1}{\sqrt{(N-1)(N-2)}}\sum_{j\neq s,r}|r,j\rangle - \sqrt{\frac{N-2}{N-1}}|r,s\rangle,\\
\nonumber |\alpha_6\rangle & = & \frac{1}{\sqrt{N-2}}\sum_{i\neq s,r}|i,s\rangle,\\
\nonumber |\alpha_7\rangle & = & \frac{1}{\sqrt{N-2}}\sum_{i\neq s,r}|i,r\rangle.
\end{eqnarray}

\begin{widetext}
The effective evolution operator is in the $|\alpha_i\rangle$ basis given by the following 7x7 matrix
$$
U_{eff} = \left(
\begin{array}{ccccccc}
 \frac{N-3}{N-1} & -\frac{2 (N-2)}{(N-1)^2} & \frac{2 (N-3)^{3/2} \sqrt{N-2}}{(N-1)^{5/2}} & 0 & \frac{2 \sqrt{N-2}}{(N-1)^2} &  \frac{4 (N-2)^{3/2}}{(N-1)^{5/2}} & \frac{2 (N-3) \sqrt{N-2}}{(N-1)^{5/2}} \\
 -\frac{2 (N-2)}{(N-1)^2} & \frac{N-3}{N-1} & \frac{2 (N-3)^{3/2} \sqrt{N-2}}{(N-1)^{5/2}} & \frac{2 \sqrt{N-2}}{(N-1)^2} & 0 &  \frac{2 (N-3) \sqrt{N-2}}{(N-1)^{5/2}} & \frac{4 (N-2)^{3/2}}{(N-1)^{5/2}} \\
 -2 \sqrt{\frac{(N-3) (N-2)}{(N-1)^3}} & -2 \sqrt{\frac{(N-3) (N-2)}{(N-1)^3}} & \frac{(N-5)^2}{(N-1)^2} & \frac{2 \sqrt{N-3}}{(N-1)^{3/2}} & \frac{2 \sqrt{N-3}}{(N-1)^{3/2}} &  \frac{2 (N-5) \sqrt{N-3}}{(N-1)^2} & \frac{2 (N-5) \sqrt{N-3}}{(N-1)^2} \\
 0 & -\frac{2 \sqrt{N-2}}{(N-1)^2} & -\frac{2 \sqrt{N-3} (N+1)}{(N-1)^{5/2}} & -\frac{N-3}{N-1} & \frac{2}{(N-1)^2} &  -\frac{4}{(N-1)^{5/2}} & \frac{2 (N-3) N}{(N-1)^{5/2}} \\
 -\frac{2 \sqrt{N-2}}{(N-1)^2} & 0 & -\frac{2 \sqrt{N-3} (N+1)}{(N-1)^{5/2}} & \frac{2}{(N-1)^2} & -\frac{N-3}{N-1} &  \frac{2 (N-3) N}{(N-1)^{5/2}} & -\frac{4}{(N-1)^{5/2}} \\
 0 & -2 \sqrt{\frac{N-2}{(N-1)^3}} & \frac{2 (N-3)^{3/2}}{(N-1)^2} & 0 & -\frac{2 (N-2)}{(N-1)^{3/2}} &  -\frac{(N-3)^2}{(N-1)^2} & \frac{2 (N-3)}{(N-1)^2} \\
 -2 \sqrt{\frac{N-2}{(N-1)^3}} & 0 & \frac{2 (N-3)^{3/2}}{(N-1)^2} & -\frac{2 (N-2)}{(N-1)^{3/2}} & 0 & \frac{2 (N-3)}{(N-1)^2} & -\frac{(N-3)^2}{(N-1)^2} \\
\end{array}
\right).
$$
Direct diagonalization of $U_{eff}$ is rather difficult, however, the eigenvalues can be determined analytically. Indeed, the characteristic equation
$$
{\rm det}\left(U_{eff} - e^{i\omega}I_7\right) = 0
$$
can be written in the form
$$
\left(5+N(N-4)+(N-1)^2\cos\omega\right)\left(-N^2 + 8N - 17  + 2(N-4)\cos{\omega} + (N-1)^2\cos^2{\omega}\right) \sin\left({\frac{\omega}{2}}\right)= 0.
$$
\end{widetext}
We find the solutions
\begin{eqnarray}
\label{omega:szegedy}
\nonumber \omega_0 & = & 0,\\
\nonumber \omega_1 & = & \arccos\left(\frac{4-N + \Delta}{(N-1)^2}\right), \\
\nonumber \omega_2 & = & \arccos\left(\frac{4-N - \Delta}{(N-1)^2}\right), \\
\omega_3 & = & \arccos\left(\frac{4N - N^2 - 5}{(N-1)^2}\right),
\end{eqnarray}
where $\Delta$ is given by
$$
\Delta = \sqrt{N^4 - 10N^3 + 35N^2 - 50N + 33}.
$$
The spectrum of the effective evolution operator $U_{eff}$ is then given by
\begin{eqnarray}
\nonumber \lambda_0 & = & e^{i\omega_0} = 1,\\
\nonumber \lambda_{1}^\pm & = & e^{\pm i\omega_1},\\
\nonumber \lambda_{2}^\pm & = & e^{\pm i\omega_2},\\
\nonumber \lambda_{3}^\pm & = & e^{\pm i\omega_3}.
\end{eqnarray}
The eigenvector corresponding to the eigenvalue $\lambda_0 = 1$ can be also determined analytically. We find that it reads
\begin{eqnarray}
\nonumber |\chi_0\rangle & = & \frac{1}{\sqrt{2}}\sqrt{\frac{N(N-3)+2}{N(N-3)+3}}(|\alpha_1\rangle - |\alpha_2\rangle) + \\
\nonumber & & + \frac{1}{\sqrt{2(N(N-3)+3)}}(|\alpha_6\rangle - |\alpha_7\rangle).
\end{eqnarray}
We point out that this eigenvector has a large overlap with the initial state of the walk $|\alpha_1\rangle$ and the desired target state $|\alpha_2\rangle$. Indeed, for large $N$ we can write
\begin{equation}
\label{szegedy:evec:0}
|\chi_0\rangle = \frac{1}{\sqrt{2}}(|\alpha_1\rangle - |\alpha_2\rangle) + O(N^{-1}).
\end{equation}
Notice that for $N\rightarrow\infty$  the vector (\ref{szegedy:evec:0}) has the same shape as the eigenvector of the walk on the star graph (\ref{star:evec:0}) corresponding to the eigenvalue $\lambda = 1$.

The explicit form of the eigenvectors $|\chi_i^\pm\rangle$ is quite lengthy. However, it turns out that for large $N$ only $|\chi_1^+\rangle$ and $|\chi_1^-\rangle$, i.e. the eigenvectors corresponding to $\lambda_{1}^\pm = e^{\pm i\omega_1}$, are relevant, since the overlaps of $|\alpha_{1,2}\rangle$ with $|\chi_j^\pm\rangle$ vanishes as $O(N^{-\frac{1}{2}})$ for $j=2,3$. We find that for large $N$ the eigenvectors $|\chi_{1}^\pm\rangle$ are given by
\begin{equation}
\label{szegedy:evec:pm}
|\chi_1^\pm\rangle = \frac{1}{2}(|\alpha_1\rangle + |\alpha_2\rangle) \pm \frac{i}{\sqrt{2}}|\alpha_3\rangle + O(N^{-\frac{1}{2}}).
\end{equation}
Again, for $N\rightarrow\infty$ the eigenvectors (\ref{szegedy:evec:pm}) have the same shape as the eigenvectors of the walk on the star graph (\ref{star:evec:pm}). Moreover, we find that the phase $\omega_1$ (\ref{omega:szegedy}) approaches (\ref{omega:star}) as $N$ tends to infinity, i.e. also the corresponding eigenvalues coincides with those for the star graph. Hence, in the limit of large $N$ the dynamics of the Szegedy's walk with queries on the complete graph reduces to the dynamics of the coined walk on the star graph. Since we have shown in Section~\ref{sec2} that the latter model achieves perfect state transfer, the same applies to the former, however, only in the limit of large $N$. We conclude that the Szegedy's walk with queries on the complete graph achieves almost perfect state transfer between the sender and the receiver vertex when we choose the number of steps $T$ as the closest integer to $\frac{\pi}{\omega_1}$, i.e.
\begin{equation}
\label{T:szegedy}
T \approx \frac{\pi}{\arccos\left(\frac{4-N + \Delta}{(N-1)^2}\right)},
\end{equation}
which approaches half the value (\ref{T:star}) required for the star graph and complete graph with self-loops as $N$ tends to infinity.

\begin{figure}[h]
\begin{center}
\includegraphics[width=0.45\textwidth]{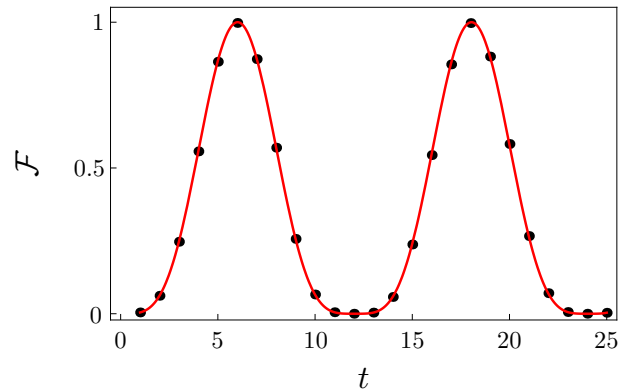}\hfill
\end{center}
\caption{Fidelity between for the Szegedy's walk with queries on the complete graph as a function of the number of steps $t$. The black dots correspond to the numerical simulation and the red line is given by (\ref{fidelity:szegedy}). We have considered the complete graph with $N=30$ vertices. The first maximum of the fidelity is reached after 6 steps, in accordance with (\ref{T:szegedy}).}
\label{fig3}
\end{figure}

For illustration we display in Figure~\ref{fig3} the fidelity between the state of the walk and the target state $|\alpha_2\rangle$ for the Szegedy's walk with queries on the complete graph with $N=30$ vertices. Within the approximations made in (\ref{szegedy:evec:0}), (\ref{szegedy:evec:pm}) the fidelity is given by
\begin{equation}
\label{fidelity:szegedy}
{\cal F}(t) = \left|\langle\psi(t)|\alpha_2\rangle\right|^2 \approx \sin^4\left(\frac{\omega_1 t}{2}\right).
\end{equation}
For the complete graph with $N=30$ vertices the first maximum of fidelity is reached after 6 steps of the walk, in agreement with the analytical prediction of (\ref{T:szegedy}).

\section{Conclusions}
\label{sec5}

State transfer between two vertices of a graph by means of discrete-time quantum walk search algorithm with two marked vertices was analyzed. In particular, we have shown that the coined quantum walk on a star graph and complete graph with self-loops achieve perfect state transfer between the sender and receiver vertex for arbitrary number of vertices $N$. On the other hand, Szegedy's walk with queries on complete graph achieves perfect state transfer only in the limit of large $N$. All three algorithms require $O(\sqrt{N})$ steps.

The present model does not allow for the transfer of the internal coin state of the particle which is possible in other discrete-time models \cite{wojcik:qw:pst,zhan:qw:pst,gedik:qw:pst}. Indeed, there is either no non-trivial internal state as for the walk on the star graph, or it has to be fixed as for the coined walk on the complete graph with self-loops and Szegedy's walk on the complete graph. On the other hand, the present method requires less control over the system, since we only have to adjust the coin at the sender and receiver vertex.

It is of interest to determine additional graphs where perfect state transfer is possible by means of discrete-time quantum walks. Our preliminary numerical analysis indicates that the modification of the Szegedy's walk where the receiver vertex is in the copy of the original graph also achieves state transfer with high fidelity. This result suggests that discrete-time quantum walks are suitable for perfect state transfer on complete bipartite graphs. We plan to thoroughly investigate this model in the near future.

Finally, let us point out that in the continuous-time quantum walk scenario Chakraborty et al. \cite{shantanav:search} have shown that state transfer with fidelity approaching unity is achieved for almost all graphs in the limit of large number of vertices $N$. It would be interesting to prove similar statement in the discrete-time case. Moreover, Chakraborty et al. \cite{shantanav:search} have also considered entanglement generation between two vertices. The protocol uses a non-adjacent third party vertex, which has to tune its nearest neighbor couplings. We plan to identify the discrete-time counterpart of this protocol.


\begin{acknowledgments}

We appreciate the financial support from RVO~68407700 and from Czech Technical University in Prague under Grant No. SGS16/241/OHK4/3T/14. M\v S is
grateful for the financial support from GA\v CR under Grant No.
14-02901P.

\end{acknowledgments}


\end{document}